\documentclass[12pt]{amsart}
\usepackage[margin=1in]{geometry}
\usepackage{mathptmx}
\usepackage{mathtools}
\usepackage{amsthm}
\usepackage{amssymb}
\usepackage[alphabetic]{amsrefs}
\usepackage{graphicx}
\usepackage{tikz}
	\usetikzlibrary{decorations.pathreplacing}
	\usetikzlibrary{decorations.pathmorphing}
	\usetikzlibrary{patterns}
\usepackage{caption}

\newcommand{\cd}{\cdot}
\newcommand{\ra}{\rightarrow}
\newcommand{\pr}{\prime}
\newcommand{\de}{\partial}

\theoremstyle{definition}
\newtheorem{defin}{Definition}
\newtheorem*{defin*}{Definition}
\theoremstyle{remark}

\begin{document}

\title{Spinoza, Leibniz, Kant, and Weyl}

\author{Michael H. Freedman}
\address{\hskip-\parindent
	Michael H. Freedman \\
	Microsoft Research, Station Q, and Department of Mathematics \\
	University of California, Santa Barbara \\
	Santa Barbara, CA 93106
}

\maketitle

I walk into a House of Worship and the guy up front is in the middle of an oration. I don't expect to stay but catch a few words, ``... so my friends, reach down and uplift thy less fortunate neighbor---to the $\frac{1}{2}$-power, for so it is written, and so may it be derived.'' Startled, I take a seat; this sounds like theology on a wavelength I can receive. This was my experience meeting (virtually) Glen Weyl and reading his paper with Buterin and Hitzig \cite{bhw}. The paper describes a funding mechanism called \emph{Quadratic Finance} (QF) and deploys a bit of calculus to show that within a very clean and simple linear model QF maximizes social utility. They differentiate the social utility function. The mathematical content of this note is that by taking one further derivative, one may also deduce that QF is the \emph{unique} solution. But what drew me into this house of worship, is a line of reasoning in [BHW], made explicit in a phone call with Glen, that Kant's \emph{categorical imperative} (CI), properly interpreted, becomes a differential equation uniquely solved by QF. It was this second derivation from Kant's ``axiom'' that caused me to take a seat, and for a very personal reason.

My father, Benedict Freedman \cites{bf1,bf2}, was a novelist, screenwriter, and educator, trained in philosophy and mathematical logic. He lived in the realm spanning Spinoza to G\"{o}del; he sought, all his life, to expand upon Spinoza's Ethics, an astonishingly ambitious project to logically deduce how to act in our world from a minimal axiomatic foundation of core beliefs. As psychologists know, beliefs are generally context dependent and fuzzy. What kind of logical machinery can they survive? Is it realistic to deduce ethics? One thought (I hope to develop later), which I believe my father would endorse, is to use a logic with a web-like rather than a linear structure, a pluralist's logic, a logic which is able to weigh preponderance of evidence from various pathways much as Feynmann diagrams in perturbative quantum field theory must be summed over to reach a reliable conclusion.

As I learned from Jerry Muller \cite{muller}, Western political and social structure, both in theory and practice, can be viewed as a secularization of Judeo-Christian religious tradition. Our quest for understanding has this broader context. This view may also be applied to the sciences, and comfortably so, in the case of economics. Marxism is a prime exemplar, with its emphasis on a goal-directed, teleological, historiography replete with catastrophism and a utopian eschatology. My father once said, ``All the great Jews of history were atheists: Spinoza, Marx, Einstein.'' He added that these three each demonstrated deep religious sensibilities. (My father, a modest man, merely considered himself a Jewish agnostic. When I asked him as a teenager how a Jew could be an atheist, he answered that atheism did not come from nowhere---it was invented by Jews, it is a sect within Judaism.) Spinoza's religious perspective encouraged his belief that there are anchor points strong enough to sustain deduction. Alas, I was never able to follow Spinoza's deductions, a shortcoming, I suspect, of a too narrowly mathematical mind, always demanding clarity of definition. For fifty years I gave all this little thought, until seeing last month, in the deduction of QF from CI, a first glimmer of success for Spinoza's (and my father's) project. That is, a success that I could understand. Imagining the future, Leibniz wrote, ``When there are disputes among persons, we can simply say, `Let us calculate', and without further ado, see who is right.'' In Leibniz's future, mathematics may usefully manipulate ethical propositions. I had not encountered a convincing example of this prior to CI $\ra$ QF. Is this future closer than we thought? New tools open new domains: Can an economic perspective, certainly foreign to Spinoza and Leibniz, lead to an ethical calculus?

Kant's CI is more subtle than the much remembered, ``A deed which must be done though the heavens fall.'' Kant wrote in \emph{Groundwork of the Metaphysics of Morals}, ``Act only according to that maxim whereby you can at the same time, will that it should become universal law.'' This rejection of exceptionalism clearly has roots in the Golden Rule but is now couched unconditionally in a more abstract form. All those interested in distributed architectures and shared governance can feel the Kantian vibe. To accept CI requires a degree of a liberal, rationalist mind set, but superficially, it seems consistent with a wide spectrum of social and political views. But once CI is accepted, I will argue, following \cite{bhw}, that it implies QF, an apparently more generous doctrine which in the orator's words, uplifts the weaker players according the $\frac{1}{2}$-power. The surprise is that a calculation is required---Leibniz take note---to learn the moral scope of one's own maxim. Curiously, the conclusion is fine-tuned, one is not lead to extremes of egalitarianism, all social differences are not leveled, but rather a certain formula for adjusting then, involving square roots and squares emerges. There is a context to the argument, standard assumptions of linearity and smoothness and concavity of individual utility functions are made, as well as the insensitivity of individual preferences to some level of taxation (to cover the shortfall between contributions and dispersed funds). Nevertheless, it is remarkable how little beyond CI is required to deduce QF. The model is simple and thus far from complete, yet seeing this deduction within the model was a moving experience, if there are other examples of ``ethical calculus,'' I would like to learn about them.

Let's review QF. QF is an answer to a question: How should public goods be financed. A ``public good,'' such as a public park, will typically be a good that is inefficient to price and restrict, and one from which individuals receive value greater than their personal contributions. Below is a brief summary of the model presented in \cite{bhw}.

\emph{Society} consists of $n$ well-defined citizens, $i = 1, \dots, n$. $p \in P$ are the public goods in question. (In our treatment we simply consider a single good, and eventually drop the $p$-index. It has no real effect on the calculus.)

Let $V_i^p(F^p)$ be the currency-equivalent utility citizen $i$ receives if the public funding of $p$ is $F^p$. These functions are assumed to be smooth, increasing and concave.\footnote{Even this assumption can often be relaxed to monotonicity.} Remarkably, nothing else about the functional form of the citizen's values needs to be assumed. We further assume all these values (utilities) are independent, i.e.\ just add, and ignore any issues of partial information and timing of decision making within the model.

$\vec{c} = \{c_i^p\}$ is the vector of individual contributions of the ith citizen toward the $p$th public good, which are all assumed non-negative (a non-trivial assumption); and $\vec{F}$ is the vector of funding with components $F^p$ for each public good. Let $C$ and $F$ denote the vector spaces holding the vectors $\vec{c}$ and $\vec{F}$.

The problem considered: Find a (and eventually all) funding mechanism $\Phi: C \ra F$ which maximizes the total social welfare $W = \sum_{i,p} V_i^p(F^p) - \sum_p F^p$.  We do not consider within the model the equity of the mechanism, which must be treated externally to the model. And very importantly we do not analyze here the back reaction on citizen contributions c which taxation inevitably entails. Such an analysis is initiated in section 4.5 of \cite{bhw}, and is shown to perturb the QF extremum, but we have no further insight to add here. Taxation $\{t_i\}$ (according to some other mechanism) is, of course, required to balance the budget:
\[
    \sum_i t_i = \sum_p\left(F^p - \sum_i c_i^p\right)
\]
and the individual's tax-corrected utility will be
\[
    U_i^t = \sum_p V_i^p(F^p) - c_i^p - t_i
\]
but I ignore taxes entirely in the calculations (as well as eventually dropping the index $p$ for visual clarity).

\subsection*{Optimality condition} Fixing $p$ and differentiating $W^p = \sum_i V_i^p(F^p) - F^p$ w.r.t.\ $F^p$ and setting this equal to zero, we find the marginal value derived from good $p$ should equal 1, provided $F^p$ is positive at 0, $\sum_i (V_i^p)^\pr = 1$. Note this sum is written ${V^p}^\pr$ in \cite{bhw}.

\subsection*{Individual utility} $U_i = \sum_p V_i^p(F^p) - c_i = \sum_p V_i^p(g(\sum_j h(c_j^p)) - c_i^p)$, where the model presumes that the functions $F^p$ are built as indicated from the internal functions $h$ and $g$, explained in the following paragraph. Below, the superscript $t$ is dropped since we are now neglecting taxes, and $p$ added to indicate the utility of a fixed good. Upon differentiating and setting $\frac{\de U_i^p}{\de c_i} = 0$, we obtain:
\begin{equation}\label{eq:indiv_util}
    \frac{\partial V_i^p}{\partial g} \cd \frac{\partial g}{\partial h(c_i^p)} \cd \frac{dh(c_i^p)}{dc_i^p} = 1
\end{equation}

In what follows differential calculations will not be interrupted to consider special case where $c_i^p = 0$, nor to reiterated the obvious boundary condition that $\Phi(0) = 0$, no contributions implies no funding.

The possible mechanisms $\Phi$ are posited to be \emph{democratic} in that the funding function $F$ is assumed to be symmetric in its $n$ variables, and beyond that I adopted the \emph{simplifying linearity assumption} of \cite{bhw} that $\Phi(\vec{c}) = \{F^P(\vec{c})\} = \{g(\sum_i h(c_i^p))\}$ for some analytic function $g$ and $h$ on the positive Reals. Our job is to find $g$ and $h$ maximizing social utility. The internal function $h$ is the \emph{weight} of contribution $c$, and $g$ converts total weight into funding, it is the \emph{funding lever}. Notice in the definition of $\Phi^{\mathrm{QF}}$ below that $h$ and $g$ scale reciprocally so funding choices are independent of the units of the currency. However, this is an outcome not an assumption.

\begin{defin}
    $\Phi^{\mathrm{QF}}(c^p) = \left(\sum_i (c_i^p)^{\frac{1}{2}}\right)^2$, that is, for every good $p$ its level of funding is the square of the sum of the square roots of the individual contributions.
\end{defin}

At this point, going forward, I drop the superscript $p$ and the set notation $\{\}$.  I will shortly set up the natural extremal problem for $\Phi$ and show that $\Phi^{\mathrm{QF}}$ is its unique critical point, but first let's show how $\Phi^{\mathrm{QF}}$ may be deduced from CI.

The insight I learned from Glen, reflected the differential equation (10) of \cite{bhw}, reproduced below, is that CI implies that if citizen $j$ deems it proper to perturb her weighted contribution $h(c_j)$, say by increasing it 1\%, she should be following, not her limited self-interest, but be justified in expecting all her peers to also see the virtue of such a similar proportional increase in their weighted contribution---``act ... whereby ... it should become universal law.'' So, mathematically we may write:
\begin{equation}\label{eq:ci}
    \frac{\partial g(\sum_i h(c_i))}{\partial c_j} = \sum_i \frac{h(c_i)}{h(c_j)}
\end{equation}

That is, the funding should respond to the imputed community wide judgement that additional matched resources are required for this good. Importantly the individual's judgement and its imputed extension must be expressed in terms of \emph{weighted} contributions, for it is the weighting function h that codifies how any given funding mechanism (obeying our assumptions) hears the preferences of its citizens.

From (\ref{eq:ci}) QF follows directly. Differentiating we obtain
\begin{equation}\label{eq:ci2}
    g^\pr \left(\sum_i h(c_i)\right) h^\pr(c_j) = \sum_i \frac{h^\pr(c_i)}{h(c_j)}
\end{equation}

Equation \ref{eq:ci2} is actually two independent equations as the factors involving only $c_j$ must separately be proportional, as must the factors involving all the $\{c_i\}$. Separating and solving one finds that for some positive constant $k$:
\begin{align}
    & g^\pr(x) = kx\ \text{ and }\ h^\pr(y) = \frac{1}{k} h^{-1}(y), \text{ so} \\
    & g(x) = \frac{k}{2} x^2 + m\ \text{ and }\ h(y) = \frac{2}{k} y^{\frac{1}{2}} + n
\end{align}

The boundary conditions tells us that $m = n = 0$, and $k$ must be set so that for a society with a single citizen, we obtain $F(c) = c$. Thus we find:
\begin{equation}
    g(x) = x^2\ \text{ and }\ h(y) = y^{\frac{1}{2}}
\end{equation}
Kant's IC has been interpreted as equation \ref{eq:ci}; which, as we have seen, has QF as its unique solution.

Let's now return now to the calculus problem of extremizing social utility $U = \sum_i(U_i)$ by varying our funding mechanism $\Phi$ though our choice of the internal functions $h$ and $g$. To do this, first on line (\ref{eq:util_max1}) we rewrite (\ref{eq:indiv_util}), and then on line (8) sum (\ref{eq:util_max1}) over $i$, apply the optimality condition, and then differentiate w.r.t.\ any one of the citizen contributions, say $c_1$ (by symmetry there is no real choice here). The result is:
\begin{align}
    & \frac{\de V_i}{\de g} = \frac{1}{\frac{\de g(\Sigma)}{\de h(c_i)} \cd \frac{d h(c_i)}{dc_i}} = \frac{1 / \frac{dh(c_i)}{dc_i}}{\frac{\de g(\Sigma)}{\de h(c_i)}},\ \text{where } \Sigma \coloneqq \sum_{i=1}^n h(c_i) = \text{ total weight} \label{eq:util_max1} \\
    & \frac{\de}{\de c_1} \left(\sum_{i=1}^n \frac{\de V_i}{\de g}\right) = \frac{\de}{\de c_1} \left(\frac{1 / \frac{dh(c_i)}{dc_i}}{\frac{\de g(\Sigma)}{\de h(c_i)}}\right) = 0
\end{align}

Now expand the outer partial using $\left(\frac{u}{v}\right)^\pr = \frac{u^\pr v - uv^\pr}{v^2}$, keeping only the numerator, and breaking out the first variable $c_1$ from the rest of the sum, we obtain:
\begin{equation}
    0 = \left(\frac{1}{h^\pr(c_1)}\right)^\pr \cd g^\pr(\Sigma) - \frac{g^{\pr\pr}(\Sigma) h^\pr(c_1)}{h^\pr(\Sigma)} + 0 - \sum_{i=2}^n \frac{1}{h^\pr(c_i)} g^{\pr\pr}(\Sigma) h^\pr(c_1)
\end{equation}

Collecting terms:
\begin{align}
    & \left(\frac{1}{h^\pr(c_1)}\right)^\pr \cd g^\pr(\Sigma) = g^{\pr\pr}(\Sigma) \left(\sum_{i=1}^n \frac{h^\pr (c_1)}{h^\pr(c_i)}\right), \text{ or} \\
    & \frac{g^\pr(\Sigma)}{g^{\pr\pr}(\Sigma)} = \frac{\sum_{i=1}^n \frac{h^\pr(c_1)}{h^\pr(c_i)}}{\left(\frac{1}{h^\pr}\right)^\pr(c_1)}, \text{ or} \\
    & (\log g^\pr)^{-1} = \frac{h^\pr(c_1)}{-\frac{h^{\pr\pr}(c_1)}{(h^\pr(c_1))^2}} \left(\sum_{i=1}^n \frac{1}{h^\pr(c_i)}\right), \text{ or} \\
    & (\log g^\pr)^{-1} = \left[- \frac{(h^\pr)^3}{h^{\pr\pr}}(c_1)\right]\left(\sum_{i=1}^n \frac{1}{h^\pr(c_i)}\right) \label{eq:log_g}
\end{align}

The factor in brackets depends only on the chosen variable, $c_1$, whereas $\log(g^\pr)(\sum_{i=1}^n \frac{1}{h^\pr(c_i)})$ is symmetric in all $n$ variables. So assuming $n > 1$, there are more than a single citizen in the society, the factor in brackets must be constant. This leads to the equation:
\begin{equation}\label{eq:brackets}
    (h^\pr)^3 = -kh^{\pr\pr}
\end{equation}
for some constant $k$.

Equation \ref{eq:brackets} may be solved recursively in a Taylor series expansion around any positive value of the variable, yielding:
\begin{equation}
    h(y) = ay^{\frac{1}{2}} + b, \text{ and } b = 0 \text{ by the boundary condition.}
\end{equation}

Thus, $\frac{1}{h^\pr(c_i)} = \frac{2}{a} y^{\frac{1}{2}}$, so $\sum_{i=1}^n \frac{1}{h^\pr(c_i)} = \frac{a^2}{2} \Sigma$, and (\ref{eq:log_g}) becomes:
\begin{equation}
    \frac{g^\pr(\Sigma)}{g^{\pr\pr}(\Sigma)} = \mathrm{const.}\ \Sigma
\end{equation}
which again by solving the Taylor series must be
\begin{equation}
    g(x) = \mathrm{const.}\ x^2 + \mathrm{const.}^\pr
\end{equation}
with $\mathrm{const.}^\pr = 0$ by the boundary condition, and const.\ $=1$ to match self-funding in the limit of one positive contribution. So we have recovered the $g$ and $h$ of QF:
\begin{equation}\label{eq:conclusion}
    g(x) = x^2 \text{ and } h(y) = y^{\frac{1}{2}}
\end{equation}

To complete the general picture of deriving QF by maximizing social utility, we can rederive the main conclusion (\ref{eq:conclusion}) more quickly by \emph{assuming} homogeneity. If one presumes in advance that $g$ and $h$ are homogeneous, $g(\Sigma) = \Sigma^q$ and $h(c) = c^p$, scale-invariance of currency implies $p = \frac{1}{q}$. Now consider the individual utility at the maximum for each $i$,
\begin{align}
    & U_i = V_i\left(\sum_j c_j^{\frac{1}{q}}\right)^q - c_i \\
    & \frac{q V_i^\pr (\sum_j c_j^{\frac{1}{q}})^{q-1}}{q c_i \frac{q-1}{q}} = 1, \text{ or } V_i^\pr = \frac{c_i^{\frac{q-1}{q}}}{\left(\sum_{j=1}^n c_j^{\frac{1}{q}}\right)^{1-q}} \text{ at criticality} \label{eq:criticality}
\end{align}

The optimality condition requires:
\begin{equation}
    1 = \sum_{i=1}^n V_i^\pr = \frac{\sum_{i=1}^n c_i^{\frac{q-1}{q}}}{\left(\sum_{j=1}^n c_j^{\frac{1}{q}}\right)^{q-1}}
\end{equation}

By H\"{o}lder's inequality, except when all $c_j$ are equal, (\ref{eq:criticality}) cannot hold except at $q=2$, again yielding
\begin{equation}
    g(x) = x^2 \text{ and } h(y) = y^{\frac{1}{2}}
\end{equation}

Indeed, there are many roads to Quadratic Finance, but Kant's categorical imperative is the most exciting---to your expositor, to Spinoza, to Leibnitz, to Kant, and to my father.

Our secular rabbi, priest, minister, and iman has done his work well and found a convert. And I doubt that the modifier \emph{secular} is even germane. As Jerry Muller has shown us, we all lie within a single broad and evolving tradition of thought. We can use both reason and love, faith for those who possess it, and the calculus of Newton and Leibnitz, for those who possess that, to navigate to a fairer, less contentious world.

\subsection*{Acknowledgment}
I thank the Aspen Center for Physics for their hospitality.

\bibliography{references}

\begin{bibdiv}
\begin{biblist}

\bib{bhw}{article}{
      author={Buterin, Vitalik},
      author={Hitzig, Zo{\"e}},
      author={Weyl, Glen},
       title={A flexible design for funding public goods},
        date={2019},
     journal={Manag. Sci.},
      volume={65},
      number={11},
       pages={5171\ndash 5187},
}

\bib{bf2}{book}{
      author={Freedman, Benedict},
      author={Freedman, Nancy},
       title={The {A}pprentice {B}astard},
   publisher={Trident Press},
        date={1966},
}

\bib{bf1}{book}{
      author={Freedman, Benedict},
       title={Rescuing {T}he {F}uture},
   publisher={Createspace Independent Pub},
        date={2010},
}

\bib{muller}{book}{
      author={Muller, Jerry},
       title={Professor of {A}pocalypse},
    subtitle={The {M}any {L}ives of {J}acob {T}aubes},
   publisher={Princeton University Press},
        date={2022},
}

\end{biblist}
\end{bibdiv}

\end{document}